\documentclass[aps,prb,showpacs,twocolumn,amsmath,amssymb]{revtex4}
\usepackage{graphicx}
\usepackage{Jonasmacros}
\usepackage{bm}

\begin{document}
\date{\today}
\title{Formation of pure two-electron triplet states in weakly coupled quantum dots attached to ferromagnetic leads}
\author{J. Fransson}
\email{jonasf@lanl.gov}
\affiliation{Department of Materials Science and Engineering, Royal Institute of Technology (KTH), SE-100 44\ \ Stockholm, Sweden}
\affiliation{Physics Department, Uppsala University, Box 530, SE-751 21\ \ Uppsala, Sweden}

\begin{abstract}
Weakly coupled quantum dots in the Pauli spin blockade regime are considered with respect to spin-dependent transport. By attaching one half-metallic and one non-magnetic lead, the Pauli spin blockade if formed by a pure triplet state with spin moment $S_z=1$ or $-1$. Furthermore, additional spin blockade regimes emerge because of full occupation in states with opposite spin to that of the half-metallic lead.
\end{abstract}
\maketitle

Spin-dependent transport in mesoscopic system is now a well-established research field\cite{zutic2004} which has attained lots of attraction recently, both experimentally\cite{tsukagoshi1999,orgassa2001,deshmukh2002,zhao2002,kim2002,petta2004} and theoretically.\cite{rudzinski2001,fransson2002,konig2003,choi2004} One major reason is the possibilities of using, for example, quantum dots (QDs) in spin-qubit readout technologies,\cite{bandyopadhyay2003}; however, also fundamental questions concerning how spin coherence and interactions influence the transport are of main interest. Recent studies on weakly coupled QDs\cite{ono2002,rogge2004,johnson2005} revealed blockaded transport due to a unit population of the two-electron triplet states, the so-called Pauli spin blockade (PSB), which has also been theoretically verified.\cite{liu2005,franssoncm2005}

In serially coupled DQDs, the PSB can be found when the energy separation of the single electron levels between the two QDs exceeds intra-dot charging energy subtracted by the inter-dot charging energy. The thus imposed inversion asymmetry on the DQD, i.e. the level off-set between the single electron levels in the two QDs, creates a unit probability to populate the two-electron triplet states in one direction of the bias voltage, henceforth referred to as \emph{forward} biases. Because of the Pauli exclusion principle, there cannot flow any electrons through the DQD once the triplet state is fully populates, hence, the systems enters the PSB.

In this paper, this system tuned into the PSB regime, where it is investigated with respect to spin-dependent transport. Replacing one of the leads by a ferromagnet, does not lift the unit probability of the population in the triplet state for forward biases. For definiteness, let forward biases mean that the chemical potential of the left lead is higher than that of the right lead. Thus, electrons flow from the left to the right lead, via the double quantum dot (DQD). Further, assume that the left lead is ferromagnetic. The broken spin symmetry creates a difference of the population probabilities for the three components of the triplet state which generates a finite spin moment in the DQD in the same direction as the majority spin in the ferromagnetic lead. Particularly, in the limiting case where the ferromagnetic lead is a half-metal with spin $\sigma\ (\sigma=\up,\down)$, the triplet is found to be in the pure two-electron configuration $\ket{\sigma}_A\ket{\sigma}_B$ with unit probability, where the subscripts indicate QD$_{A/B}$, where QD$_A$ (QD$_B$) is adjacent to the left (right) lead. This can be understood since the triplet consists of three states, 
hence, there is always at least one of the three states available for occupation independently of the spin-polarisation in the leads. Since the sum of population probabilities for these states is unity in the blockade regime, the population is distributed among the states in the most probable combination.

For reverse biases the system behaves somewhat differently when one of the non-magnetic leads is replaced by a half-metal. In the non-magnetic case,\cite{ono2002,franssoncm2005} the current is mediated through transitions between the two lowest singlet states and the lowest (spin-degenerate) one-electron state. The probabilities for these transitions are about 1/2 to both leads. In the magnetic case with the left lead half-metallic (right lead non-magnetic), however, only electrons of the same spin as those in the half-metallic left lead are allowed to proceed through the DQD. Since both spin directions are available in the right lead and electrons of both spins are equally probable to tunnel into the DQD, the flow of electrons from the right tends to accumulate electrons in the state $\ket{\bar\sigma}_A\ket{\bar\sigma}_B$, that is, opposite to the spin in the half-metal. This accumulation is provided by tunneling processes like, for example, $\ket{\text{initial}}=\ket{0}_A\ket{\bar\sigma}_B\ \rightarrow\ \ket{\bar\sigma}_A\ket{0}_B\ \rightarrow\ \ket{\bar\sigma}_A\ket{\bar\sigma}_B=\ket{\text{final}}$, where it is understood that the electrons flow from the right to the left lead.


If, instead, the right lead is half-metallic (left lead non-magnetic), this configuration also causes a blockaded current for forward biases. However, in this case the blockade is caused by a unit population probability of the  triplet state $\ket{\bar\sigma}_A\ket{\bar\sigma}_B$, that is, the one with opposite spin to that of the half-metal. The reason is clear since electrons with the same spin as the half-metal can tunnel through the DQD whereas those of opposite spin are blockaded by the absence of such spin states in the right lead. For reverse biases, the flowing electrons only have one spin direction (that of the half-metallic right lead). Since the singlet states require two electrons with opposite spin, these are unavailable for population, hence, for transport. In addition, since the probability for population in the triplet state is negligible when the left lead is non-magnetic,\cite{franssoncm2005} only the one-electron state with the same spin as the half-metallic right lead is populated. This too leads to a suppressed current.

The discussion is illustrated by considering two single level QDs $(\dote{A},\dote{B}$, spin-degenerate) with intradot charging energies $(U_A=U_B=U)$, which are coupled by interdot charging $(U')$ and tunnelling $(t)$ interactions, or more specifically\cite{sandalov1995,inoshita2003,cota2005,franssoncm2005} $\Hamil_{DQD}=\sum_{i=A,B}(\sum_\sigma\dote{i}\ddagger{i\sigma}\dc{i\sigma}+U_in_{i\up}n_{i\down})+U'(n_{A\up}+n_{A\down})(n_{B\up}+n_{B\down})+\sum_\sigma(t\ddagger{A\sigma}\dc{B\sigma}+H.c.)$. Here $\ddagger{A\sigma/B\sigma}\ (\dc{A\sigma/B\sigma})$ creates (annihilates) an electron in QD$_{a/B}$ at the energy $\dote{A\sigma/B\sigma}$, while $n_{A\sigma}=\ddagger{A\sigma}\dc{A\sigma}$ and analogous for $n_{B\sigma}$. The PSB regime is obtained for $\dote{A}-\dote{B}=\Delta\dote{}>0$, $\Delta\dote{}<2(\mu-\dote{A})<5\Delta\dote{}$, where $\mu$ is the equilibrium chemical potential, $U=2U'=2\Delta\dote{}$, and weakly coupled QDs, e.g. $\xi=2t/\Delta\dote{}\ll1$. The condition $\Delta\dote{}>0$ imposes the required inversion asymmetry on the system ($\Delta\dote{}<0$ provides the PSB for reversed biases). As shown in Ref. \onlinecite{franssoncm2005}, the range of chemical potentials satisfying $2(\mu-\dote{A})\in(\Delta\dote{},5\Delta\dote{})$ provides a regime where the transport is mediated between one- and two-electron states in which the PSB is available. Generally also low temperatures are required, e.g. $k_BT/U\leq0.025$. For a more detailed discussion on the mechanism behind the PSB regime for this system and the choice of parameters, as well as for the model of the DQD coupled to leads, the reader is referred to Ref. \onlinecite{franssoncm2005}.

This configuration of the DQD and biases voltages $V\in(-1,1)U/e$, enables the analysis performed in Ref. \onlinecite{franssoncm2005} (to which the reader is referred for the general equations and notation). Briefly, a density matrix approach for the population number probabilities of the many-body eigenstates have been used. The DQD has 16 eigenstates with corresponding population probabilities $P_{Nn}$, denoting the $n$th population probability of the $N$-electron configuration ($N=0,\ldots,4$). In the Markovian approximation (which is sufficient for stationary processes) the equations for the population probabilities $P_{Nn}$, to the first order in the couplings $\Gamma^{L/R}_\sigma=2\pi\sum_{k\in L/R}|v_{k\sigma}|^2\delta(\omega-\leade{k})$ to the left/right $(L/R)$ lead, can be written as
\begin{widetext}
\begin{eqnarray}
\ddt P_{Nn}&=&
\frac{1}{\hbar}\sum_{\chi=L,R}\biggl(\sum_m\Gamma^\chi_{N-1m,Nn}
	[f^+_\chi(\Delta_{Nn,N-1m})P_{N-1m}
	-f^-_\chi(\Delta_{Nn,N-1m})P_{Nn}]
\nonumber\\&&
	-\sum_m\Gamma^\chi_{Nn,N+1m}
	[f^+_\chi(\Delta_{N+1m,Nn})P_{Nn}
	-f^-_\chi(\Delta_{N+1m,Nn})P_{N+1m}]
	\biggr)=0,
\label{eq-P}\\
N&=&0,\ldots,4,
\nonumber
\end{eqnarray}
\end{widetext}
where $P_{-1n}=P_{5n}\equiv0$. Here, $\Delta_{N+1m,Nn}$ denote the energies for the transitions $\ket{N,n}\bra{N+1,m}$, while $\Gamma^{L/R}_{Nn,N+1m}=\sum_\sigma\Gamma^{L/R}_\sigma(\dc{A\sigma/B\sigma})^{nm}_{NN+1}$, where $(\dc{A\sigma/B\sigma})^{nm}_{NN+1}$ is the transition matrix element for an electron exiting QD$_{A/B}$. Moreover, $f^+_\chi(\omega)=f(\omega-\mu_\chi)$ is the Fermi function at the chemical potential $\mu_\chi$ of the lead $\chi=L,R$, whereas $f^-_\chi(\omega)=1-f^+_\chi(\omega)$. Effects from off-diagonal occupation numbers, which only appear in the second order (and higher) in the couplings, are neglected since these include off-diagonal transition matrix elements to the second order (and higher) which generally are small for $\xi\ll1$.

Ferromagnetic leads are introduced through the parametrised couplings\cite{fransson2005}  $\Gamma^L_{\up,\down}=\Gamma_0(1\pm p_L)/2$ and $\Gamma^R_{\up,\down}=\Gamma_0(1\pm p_R)/2$. Here, $p_{L/R}\in[-1,1]$ accounts for the magnetisation in the left/right lead such that $p_\chi=0$ corresponds to a non-magnetic lead while $p_\chi=1\ (-1)$ corresponds to a half-metallic lead with spin $\up\ (\down)$.

In the given bias voltage regime, it turns out that the only non-vanishing population probabilities\cite{franssoncm2005} are $P_{1m},\ m=1,2$, for the one-electron states $\ket{1,m}=\alpha\ket{\sigma_m}_A\ket{0}_B+\beta\ket{0}_A\ket{\sigma_m}_B$, $\sigma_{1,2}=\up,\down$, and $P_{2n},\ n=1,\ldots,5$, for the two-electron states $\ket{2,n}=\ket{\sigma_n}_A\ket{\sigma_n}_B,\ n=1,2$, $\sigma_{1,2}=\up,\down$, $\ket{2,3}=[\ket{\up}_A\ket{\down}_B+\ket{\down}_A\ket{\up}_B]/\sqrt{2}$, and $\ket{2,n}=A_n[\ket{\up}_A\ket{\down}_B-\ket{\down}_A\ket{\up}_B]/\sqrt{2}+B_n\ket{\up\down}_A\ket{0}+C_n\ket{0}_A\ket{\up\down}_B,\ n=4,5$. Here, the two-electron eigenenergies are related as $E_{24}\leq E_{2n}\leq E_{25},\ n=1,2,3$. Here also, $\alpha,\ \beta,\ A_n,\ B_n$, and $C_n$ are coefficients, which depend on the internal parameters ($\dote{A},\dote{B},t,U',U$) of the DQD, for the eigenstates expanded in the Fock basis.

First consider forward biases in the range\cite{franssoncm2005} $eV\in(0.1,1)U$. Then,  the non-equilibrium conditions yield $f_L(\Delta_{2n,1m})\approx1$ and $f_R(\Delta_{2n,1m})\approx0$, where $f_{L/R}(x)=f(x-\mu_{L/R})$ is the Fermi function at the chemical potential $\mu_{L/R}$, whereas $\Delta_{2n,1m}=E_{2n}-E_{1m}$ is the energy for the transition $\ket{1,m}\bra{2,n}$. This leads to the following equations for the population probabilities:
\begin{subequations}
\label{eq-posP1}
\begin{eqnarray}
P_{1m}&=&\frac{\sum_{\sigma,n=1}^5
			\Gamma^R_\sigma|(\dc{B\sigma})_{12}^{mn}|^2P_{2n}}
		{\sum_{\sigma,n=1}^5\Gamma^L_\sigma|(\dc{A\sigma})_{12}^{mn}|^2},
		\ m=1,2,
\label{eq-P1}\\
P_{2n}&=&\frac{\sum_{\sigma,m=1}^2
			\Gamma^L_\sigma|(\dc{A\sigma})_{12}^{mn}|^2P_{1m}}
		{\sum_{\sigma,m=1}^2\Gamma^R_\sigma|(\dc{B\sigma})_{12}^{mn}|^2},
		\ n=1,\ldots,5.
\label{eq-P2}
\end{eqnarray}
\end{subequations}
These equations are further simplified by using the transition matrix elements $|(\dc{A\sigma/B\sigma})_{12}^{mn}|^2$ in Table \ref{tab-mte}. A somewhat tedious but straightforward calculation yields
\begin{subequations}
\label{eq-posP2}
\begin{eqnarray}
P_{11}&=&\frac{1+p_R}{1+p_L}\biggl(\frac{\alpha}{\beta}\biggr)^2P_{21},
\label{eq-P11}\\
P_{12}&=&\frac{1-p_L}{1-p_R}\biggl(\frac{1+p_R}{1+p_L}\biggr)^2
	\biggl(\frac{\alpha}{\beta}\biggr)^2P_{21},
\label{eq-P12}\\
P_{22}&=&\biggl(\frac{1-p_L}{1+p_L}\frac{1+p_R}{1-p_R}\biggr)^2P_{21},
\label{eq-P22}\\
P_{23}&=&\frac{1-p_L}{1+p_L}\frac{1+p_R}{1-p_R}P_{21},
\label{eq-P23}\\
P_{2n}&=&\frac{1-p_L}{1+p_L}\frac{1+p_R}{1-p_R}
	\biggl(\frac{\alpha}{\beta}\frac{L_n}{R_n}\biggr)^2P_{21},
\label{eq-P2n}
\end{eqnarray}
\end{subequations}
where $L_n=\beta A_n/\sqrt{2}+\alpha B_n$ and $R_n=\alpha A_n/\sqrt{2}+\beta C_n$. Conservation of charge here implies that $1=\sum_1^2P_{1m}+\sum_1^5P_{2n}$, which gives
\begin{eqnarray}
P_{21}&=&\biggl\{1+\frac{1+p_R}{1+p_L}\biggl[\frac{1-p_L}{1-p_R}
	+\frac{1+p_R}{1+p_L}\biggl(\frac{1-p_L}{1-p_R}\biggr)^2
\nonumber\\&&
	+\biggl(\frac{\alpha}{\beta}\biggr)^2\biggl(
		1+\frac{1-p_L}{1-p_R}\frac{1+p_R}{1+p_L}
\nonumber\\&&
		+\frac{1-p_L}{1-p_R}\sum_{4,5}
			\biggl(\frac{L_n}{R_n}\biggr)^2\biggr)\biggr]\biggr\}^{-1}.
\label{eq-P21pos}
\end{eqnarray}
\begin{table}[t]
\caption{Matrix elements for the relevant transitions expressed in terms of the eigenvector coefficients. Here, $L_n^2=(\beta A_n/\sqrt{2}+\alpha B_n)^2$ and $R_n^2=(\alpha A_n/\sqrt{2}+\beta C_n)^2$.}
\label{tab-mte}
\begin{center}
\begin{tabular}{l}
\hline
$|(\dc{A\up})^{11}_{12}|^2=|(\dc{A\down})^{22}_{12}|^2=\beta^2$\\
$|(\dc{A\up})^{12}_{12}|^2=|(\dc{A\down})^{21}_{12}|^2=0$\\ $|(\dc{A\down})^{13}_{12}|^2=|(\dc{A\up})^{23}_{12}|^2=\beta^2/2$\\ $|(\dc{A\down})^{1n}_{12}|^2=|(\dc{A\up})^{2_n}_{12}|^2
	=L_n^2,\ n=4,5$\\
$|(\dc{B\up})^{11}_{12}|^2=|(\dc{B\down})^{22}_{12}|^2=\alpha^2$\\ $|(\dc{B\up})^{12}_{12}|^2=|(\dc{B\down})^{21}_{12}|^2=0$\\ $|(\dc{B\down})^{13}_{12}|^2=|(\dc{B\up})^{23}_{12}|^2=\alpha^2/2$\\ $|(\dc{B\down})^{1n}_{12}|^2=|(\dc{B\up})^{2n}_{12}|^2
	=R_n^2,\ n=4,5$\\
\hline
\end{tabular}
\end{center}
\end{table}

Now, consider the case $p_1=1$ and $p_R=0$, which corresponds to letting the left lead be half-metallic with spin $\up$ whereas the right is non-magnetic. In this case the population number $P_{21}$ 
reduces to $P_{21}=1/[1+(\alpha/\beta)^2/2]$, where $\alpha^2=\xi^2/[(1+\sqrt{1+\xi^2})^2+\xi^2]$ and $\beta^2=(1+\xi^2)/[(1+\sqrt{1+\xi^2})^2+\xi^2]$ giving $(\alpha/\beta)^2=\xi^2/(1+\xi^2)$. Hence, 
\begin{equation}
P_{21}=2\frac{1+\xi^2}{2+3\xi^2}\rightarrow1\ \text{as}\ \xi\rightarrow0.
\label{eq-posPSB}
\end{equation}
Physically, this results mean that the population probability of the triplet configuration $\ket{\up}_A\ket{\up}_B$ is unity. Consistency with Eq. (\ref{eq-posP2}) is immediate, since all but Eq. (\ref{eq-P11}) vanish for $p_L=1$, while $P_{11}\propto(\alpha/\beta)^2\rightarrow0,\ \xi\rightarrow0$. The result is also consistent with the normal PSB, since the population probability of the triplet states must be unity.  Hence, while only spin $\up$ electrons can tunnel from the left leads into the DQD, the only configuration available for these electrons is $\ket{\up}_A\ket{\up}_B$. 

Similarly, let $p_L=0$ and $p_R=-1$, which corresponds to a non-magnetic left lead and half-metallic right lead with spin $\down$. Then, one finds that $P_{21}=1$ apparently independently of the coupling strength between the QDs. Thus, the state $\ket{\up}_A\ket{\up}_B$ acquires a unit population probability for a spin $\down$ half-metallic lead to the right. Consistency with Eq. (\ref{eq-posP2}) holds since all equations vanish for $p_R=-1$. A more careful analysis shows that the approximation, e.g. putting $f_L(\Delta_{2n,1m})=1$ and $f_R(\Delta_{2n,1m})=0$, provides a slightly overestimated population probability for $\ket{\up}_A\ket{\up}_B$. Nonetheless, configuring the system in this fashion indeed yields a significantly strong blockade.

By turning to the reverse biasing case, e.g. $eV\in(-1,-0.1)U$, one finds from analogous arguments, e.g. $f_L(\Delta_{2n,1m})\approx0$ and $f_R(\Delta_{2n,1m})\approx1$, that
\begin{subequations}
\label{eq-negP2}
\begin{eqnarray}
P_{11}&=&\frac{1+p_R}{1+p_L}\biggl(\frac{1-p_L}{1-p_R}\biggr)^2
	\biggl(\frac{\beta}{\alpha}\biggr)^2P_{22},
\label{eq-P11-}\\
P_{12}&=&\frac{1-p_L}{1-p_R}\biggl(\frac{\beta}{\alpha}\biggr)^2P_{22},
\label{eq-P12-}\\
P_{21}&=&\biggl(\frac{1-p_L}{1+p_L}\frac{1+p_R}{1-p_R}\biggr)^2P_{22},
\label{eq-P21-}\\
P_{23}&=&\frac{1-p_L}{1+p_L}\frac{1+p_R}{1-p_R}P_{22},
\label{eq-P23-}\\
P_{2n}&=&\frac{1-p_L}{1+p_L}\frac{1+p_R}{1-p_R}
	\biggl(\frac{\beta}{\alpha}\frac{R_n}{L_n}\biggr)^2P_{22}.
\label{eq-P2n-}
\end{eqnarray}
\end{subequations}
Conservation of charge thus implies that
\begin{eqnarray}
P_{22}&=&\biggl\{1+\frac{1-p_L}{1-p_R}\biggl[\frac{1+p_R}{1+p_L}
	+\frac{1-p_L}{1-p_R}\biggl(\frac{1+p_R}{1+p_L}\biggr)^2
\nonumber\\&&
	+\biggl(\frac{\beta}{\alpha}\biggr)^2\biggl(
		1+\frac{1-p_L}{1-p_R}\frac{1+p_R}{1+p_L}
\nonumber\\&&
		+\frac{1+p_R}{1+p_L}\sum_{4,5}
			\biggl(\frac{R_n}{L_n}\biggr)^2\biggr)\biggr]\biggr\}^{-1}.
\label{eq-P22neg}
\end{eqnarray}
It follows from Eq. (\ref{eq-P22neg}) that a half-metallic spin $\up$ left lead and non-magnetic right lead, e.g. $p_L=1$ and $p_R=0$, yields $P_{22}=1$ independently of $\xi$ since all terms in the denominator but the first vanish for this configuration. Moreover, all relations in Eq. (\ref{eq-negP2}) vanish for $p_L=1$. Physically this implies that the two-electron state $\ket{\down}_A\ket{\down}_B$ acquires a unit population probability for low temperatures whenever the left lead is a spin $\up$ half-metal.

Assuming instead that the right lead is a spin $\down$ half-metal whereas the left lead is non-magnetic, e.g. $p_L=0$ and $p_R=-1$, leads to the conclusion that $P_{22}=2\xi^2/(1+3\xi^2)\rightarrow0$ as $\xi\rightarrow0$. Further, all relations in Eq. (\ref{eq-negP2}) but Eq. (\ref{eq-P12-}) vanish in this case, whereas
\begin{equation}
P_{12}=\frac{1}{2}\biggl(\frac{\beta}{\alpha}\biggr)^2P_{22}
	=\frac{1+\xi^2}{1+3\xi^2}\rightarrow1\ \text{as}\ \xi\rightarrow0,
\label{eq-negSEB}
\end{equation}
that is, the one-electron state $\alpha\ket{\down}_A\ket{0}+\beta\ket{0}_A\ket{\down}_B$ acquires an almost full occupation for weakly coupled QDs. Hence, the DQD end up in a spin-polarized configuration, however, with only one instead of two electrons.

\begin{figure}[t]
\begin{center}
\includegraphics[width=8.5cm]{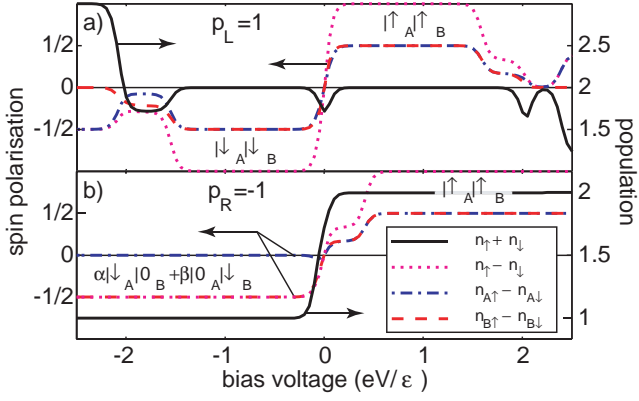}
\end{center}
\caption{(Colour online) DQD system with a) $p_L=1,\ p_R=0$, b) $p_L=0,\ p_R=-1$. Population (solid) and spin-polarisation (dotted) of the DQD as function of the bias voltage, as well as the spin-polarisations of QD$_A$ (dash-dotted) and QD$_B$ (dashed). Here, $\xi=0.01$, $k_BT/U=0.01$.}
\label{fig-NV}
\end{figure}
The plots in Fig. \ref{fig-NV} display examples of the population (solid) and spin-polarisation (dotted) in the DQD when coupled to one non-magnetic lead and one half-metallic lead a) $p_L=1,\ p_R=0$ and b) $p_L=0,\ p_R=-1$. Case a) shows regimes for both positive and negative biases in which the DQD is populated by two electrons, where the spin-polarisation is 1 and $-1$ respectively, corresponding to the cases treated in the analytical calculation above. Studying the individual spins of the QDs, QD$_A$ (dash-dotted) and QD$_B$ (dashed), shows that the two-electron regimes are indeed constituted by the triplet state $\ket{\up}_A\ket{\up}_B$ and $\ket{\down}_A\ket{\down}_B$ for positive and negative biases, respectively. In case b), there is only one two-electron regime in which the spin-polarisation is 1. Thus, the two-electron state consists of only the triplet state $\ket{\up}_A\ket{\up}_B$ for positive biases. For negative biases, the DQD is populated by a single spin $\down$ electron in the $\alpha\ket{\down}_A\ket{0}_B+\beta\ket{0}_A\ket{\down}_B$ state, as expected from the analytical analysis. The weak coupling between the QDs, i.e. $\xi\ll1$, implies that $\alpha\approx0$ and $\beta\approx1$, that is, the single spin $\down$ electron is mainly located in QD$_B$.

Although the main focus in the plots shown in Fig. \ref{fig-NV} is on the blockade regimes, there are a few details in that require some extra discussion. However, the structure for biases $|eV|/\Delta\dote{}>2$ goes beyond the focus of the present discussion, since this regime requires an additional investigation of the population numbers. In case a), the dip in the population around equilibrium ($eV=0$) is due to that the lowest two-electron singlet states (as well as the triplet) are nearly degenerate with the lowest one-electron states, hence, the energy required for transitions between those states is vanishingly small. This gives rise to an averaged population which is less than two electrons. The spin-polarization goes through zero at $eV=0$ since the lowest two-electron state is spin singlet, hence the spin-polarization vanishes. The high bias voltage dips ($eV/\Delta\dote{}\approx-1.6$ and $eV/\Delta\dote{}\approx2$) are due to the lifting of the triplet blockades caused by the availability of more transitions. In case b), the population goes from one to two electrons around equilibrium, which verifies the analytical discussion above. The spin-polarization of the DQD, however, has a regime for small positive biases where the spin $\up$ is slightly favored in both QDs but where the polarization has not saturated. This situation arises since the two electrons in the DQD is distributed among the three triplet states, although there is a slight overweight on the $\ket{\up}_A\ket{\up}_B$ state. As is seen, this state is eventually fully occupied as the bias voltage becomes sufficiently large.

\begin{figure}[t]
\begin{center}
\includegraphics[width=8.5cm]{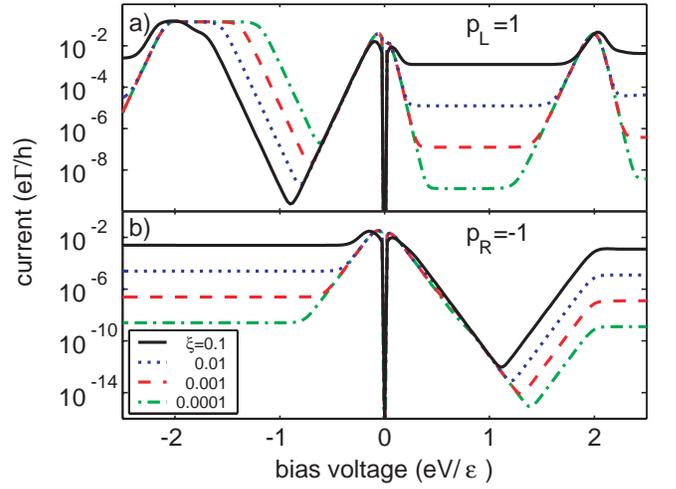}
\end{center}
\caption{(Colour online) Magnitude of the current through the system for various strengths of the coupling $\xi$ of the QDs. Other parameters as in Fig. \ref{fig-NV}.}
\label{fig-JV}
\end{figure}
The current through the system is calculated by the formula\cite{jauho1994} $J_L=-(e/h)\tr\im\int\bfGamma^L[f_L(\omega)\bfG^>(\omega)+(1-f_L)(\omega)\bfG^<(\omega)]d\omega$, where $\bfG^{</>}=\{G^{</>}_{Nm,N+1n}\}_{Nmn}$ is the matrix lesser/greater Green function (GF) of the DQD, whereas $\bfGamma^L=\sum_\sigma\Gamma^L_\sigma\{|(\dc{A\sigma})_{NN+1}^{mn}|^2\}_{Nmn}$ is the matrix of the coupling to the left lead times the matrix elements for the transitions. The lesser and greater GFs are identified by\cite{franssoncm2005} $G^>_{Nm,N+1n}(\omega)=-i2\pi P_{Nm}\delta(\omega-\Delta_{N+1n,Nm})$ and $G^<_{Nm,N+1n}(\omega)=i2\pi P_{N+1n}\delta(\omega-\Delta_{N+1n,Nm})$, respectively.

First consider case a), i.e. $p_L=1,\ p_R=0$. Then, in the forward PSB regime, i.e. $f_L(\Delta_{2n,1m})=1$, the expression for the current reduces to
\begin{equation}
J_L\approx\frac{e\Gamma_0}{\hbar}
	\frac{\alpha^2/2}{1+(\alpha/\beta)^2/2}\rightarrow0^+,\ \xi\rightarrow0.
\label{eq-JpL}
\end{equation}
Hence, for weakly coupled QDs the current is significantly suppressed, and decreases as $\xi^2,\ \xi\rightarrow0$. For negative biases the system enters the PSB regime as $f_L(\Delta_{2n,1m})=0$, giving $J_L=0$ because of the simplifications leading to Eqs. (\ref{eq-negP2}) and (\ref{eq-P22neg}). Nonetheless, the result clearly indicates a significant current suppression.

The simplified analysis is verified by the calculated current, displayed in Fig. \ref{fig-JV} a), showing the magnitude of the current as function of the bias voltage for various coupling strengths $\xi$. The plots clearly demonstrates that the forward current in the blockade regime decreases as $\xi^2$. In the reverse bias blockade, however, the current grows roughly by $\xi$ as $\xi\rightarrow0$. The reason is that electrons in the right lead enter the DQD via transitions between the triplet and one-electron states with probability $\alpha^2\sim\xi^2$. Thus, for extremely weakly coupled QDs, the triplet state acquires a non unit population probability. The remaining probability is distributed among the singlet and one-electron states, which lifts the blockade. The pronounced $V$-shape of the current for reverse biases arise from the fact that the population in the triplet state is not a completely flat function of the bias. Instead, the state $\ket{\down}_A\ket{\down}_B$ is maximally occupied around $eV/\Delta\dote{}\approx-1$ which, hence, leads to the minimal current at this bias. The dip at equilibrium is because the logarithm of the modulus of the current, i.e. $\log_{10}|J_L|$, is shown rather than the smooth function $J_L$. For higher bias voltages, e.g. $|eV|/\Delta\dote{}>2$, the triplet blockade is lifted, which is shown as a significant increase in the current (several orders of magnitude) through the DQD.

The case b), i.e. $p_L=0,\ p_R=-1$, is treated similarly. For positive the biases the above analysis provides $J_L=0$, which indicates that the current should be significantly suppressed in the forward blockade regime. For negative biases in the blockade regime, i.e. $f_L(\Delta_{2n,1m})=0$, the expression for the current reduces to
\begin{equation}
J_L\approx-\frac{e\Gamma_0}{\hbar}
	\frac{\alpha^2}
		{1+2(\alpha/\beta)^2}\rightarrow0^-,\ \xi\rightarrow0.
\label{eq-JpR}
\end{equation}
The plots in Fig. \ref{fig-JV} b) display the calculated current, showing that the forward current is dramatically reduced in the blockade regime. The blockade is strengthened by a weakened QD coupling $\xi$, which is reasonable from the analysis in the non-magnetic case.\cite{franssoncm2005} The reverse current, in the regime with a stable one-electron configuration, is also shown to be substantially suppressed (decreasing as $\xi^2$), as expected from Eq. (\ref{eq-JpR}). The $V$-shaped current for forward biases stems from the same reason as in case a).

In the present analysis a few approximations have been introduced. First, the density matrix of population probabilities is diagonal. This is reasonable since the probability for off-diagonal transitions, i.e. transitions between different states with the same number of electrons vanish in the sequential tunnelling limit. Taking this limit is justified by the large off-set $\Delta\dote{}>0$ between the QD levels.\cite{franssoncm2005} The presented results are also confirmed in models with more levels in the QDs.\cite{fransson2006} Secondly, ferromagnetic leads may induce a spin-split of the states in the DQD which is of the order of the couplings $\Gamma^{L/R}_\sigma$ between the DQD and the leads.\cite{fransson2002} Weak couplings $\Gamma^{L/R}_\sigma$, however, only marginally modifies the presented results.

The conditions required in order to obtain the PSB regime are experimentally accessible.\cite{ono2002,rogge2004,johnson2005} Half-metallic leads may be constructed from NiMnSb,\cite{groot1983} which is compatible with existing semiconductor technology.\cite{wijs2001} Another interesting half-metallic material is\cite{ji2001} Cr0$_2$ because of its favorable switching behaviour at low fields.\cite{yang2000}

\end{document}